\documentclass[10pt, conference]{IEEEtran}
\usepackage{subfigure}
\usepackage{setspace}
\usepackage{amsmath}
\usepackage{amssymb}
\usepackage{amsfonts}
\usepackage{amscd}
\usepackage{mathrsfs}
\usepackage[final]{graphicx}
\usepackage{graphicx}
\usepackage{psfrag}
\usepackage{color}
\usepackage{url}

\newtheorem{definition}{Definition}

\begin{document}

\title{A Low-Complexity, Full-Rate, Full-Diversity  $2\times2$ STBC with Golden Code's Coding Gain}

\author{
\authorblockN{K. Pavan Srinath}
\authorblockA{Dept of ECE, Indian Institute of science \\
Bangalore 560012, India\\
Email:pavan@ece.iisc.ernet.in\\
}
\and
\authorblockN{B. Sundar Rajan}
\authorblockA{Dept of ECE, Indian Institute of science \\
Bangalore 560012, India\\
Email:bsrajan@ece.iisc.ernet.in\\
}
}

\maketitle
\begin{abstract}
This paper presents a low-ML-decoding-complexity, full-rate, full-diversity  space-time block code (STBC) for  a $2$ transmit antenna, $2$ receive antenna multiple-input multiple-output (MIMO) system, with coding gain equal to that of the best and well known Golden code for any QAM constellation. Recently, two codes have been proposed (by Paredes, Gershman and Alkhansari  and by Sezginer and Sari), which enjoy a lower decoding complexity relative to the Golden code, but have lesser coding gain. The $2\times 2$ STBC presented in this paper has lesser decoding complexity  for non-square QAM constellations, compared with that of the  Golden code, while having the same decoding complexity for square QAM constellations. Compared with the Paredes-Gershman-Alkhansari and Sezginer-Sari codes, the proposed code has the same  decoding complexity for non-rectangular QAM  constellations. Simulation results, which compare the codeword error rate (CER) performance, are presented.
\end{abstract}

\section{INTRODUCTION}
Multiple-input, multiple-output(MIMO) wireless transmission systems have  been intensively studied during the last decade.  The Alamouti code \cite{SMA} for two transmit antennas is a novel scheme for MIMO transmission, which, due to its orthogonality properties, allows a low complexity maximum-likelihood (ML) decoder. This scheme led to the generalization of STBCs from orthogonal designs \cite{TJC}. Such codes allow the transmitted symbols to be decoupled from one another and single-symbol ML decoding is achieved over \emph{quasi static} Rayleigh fading channels. Even though these codes achieve the maximum diversity gain for a given number of transmit and receive antennas and for any arbitrary complex constellations, unfortunately, these codes are not $full-rate$, where, by a $full-rate$ code, we mean a code that transmits at a rate of $min(n_r,n_t)$ complex symbols per channel use for an $n_t$ transmit antenna, $n_r$ receive antenna system.


The Golden code \cite{BRV} is a full-rate, full-diversity code and has a decoding complexity of the order of $M^4,$ for arbitrary constellations of size $M.$  The codes in \cite{DV} and the trace-orthogonal cyclotomic code in \cite{ZLW} also match the Golden code. With reduction in the decoding complexity being the prime objective, two  new full-rate, full-diversity codes have recently been discovered: The first code was independently discovered by Hottinen, Tirkkonen and Wichman \cite{HTW} and by Paredes, Gershman and Alkhansari  \cite{PGA}, which we call the HTW-PGA code and the second, which we call the Sezginer-Sari code, was reported in \cite{SS} by Sezginer and Sari. Both these codes enable simplified decoding, achieving a complexity of the order of $M^3$. The first code is also shown to have the non-vanishing determinant property \cite{PGA}. However, these two codes have lesser coding gain compared to the Golden code. A detailed discussion of these codes has been made in  \cite{BHV}, wherein a comparison of the codeword error rate (CER) performance reveals that the Golden code has the best performance.

In this paper, we propose a new full-rate, full-diversity STBC for $2\times2$ MIMO transmission, which has low decoding complexity. The contributions of this paper may be summarized (see Table \ref{table1} also) as follows:
\begin{itemize}
\item The proposed code has the same coding gain as that of the Golden code (and hence of that in \cite{DV} and the trace-orthonormal cyclotomic code) for any QAM constellation (by a QAM constellation we mean any finite subset of the integer lattice) and larger coding gain than those of the HTW-PGA code and the Sezginer-Sari code.

\item Compared with the Golden code and the codes in \cite{DV} and \cite{ZLW}, the proposed code has lesser decoding complexity for all complex constellations except  for square QAM constellations in which case the complexity is the same. Compared to the HTW-PGA code and the Sezginer-Sari codes, the proposed code has the same  decoding complexity for all non-rectangular QAM [Fig \ref{fig2}] constellations.

\item The proposed code has the non-vanishing determinant property for QAM constellations
and hence is Diversity-Multiplexing Gain (DMG) tradeoff optimal.
\end{itemize}

The remaining content of the paper is organized as follows: In Section \ref{sec2}, the system model and the code design criteria are reviewed along with some basic definitions. The proposed STBC is described in Section \ref{sec3} and its non-vanishing determinant property is shown in  Section \ref{sec4}. In Section \ref{sec5} the ML decoding complexity of the proposed code is discussed and the scheme to decode it using sphere decoding is discussed in Section \ref{sec6}. In Section \ref{sec7}, simulation results are presented to show the performance of the proposed code as well as to compare with few other known codes. Concluding remarks constitute Section \ref{sec8}.

\textit{Notations:} For a complex matrix  $X,$ the matrices $X^T$, $X^{H}$ and $det\left[X\right]$ denote the transpose, Hermitian and determinant of $X,$ respectively. For a complex number $s,$ $\mathcal{R}\left(s\right)$ and $\mathcal{I}\left(s\right)$ denote the real and imaginary part of $s,$ respectively. Also, $j$ represents $\sqrt{-1}$ and the set of all integers, all real and complex numbers are denoted by $\mathbb{Z},$  $\mathbb{R}$ and $\mathbb{C},$ respectively. The Frobenius norm and the trace are denoted by $\Vert.\Vert_F$ and $tr\left[.\right] $ respectively. The columnwise stacking operation on $X$ is denoted by $vec(X).$ The Kronecker product is denoted by $\otimes$ and $I_{T}$ denotes the $T\times T$ identity matrix. Given a complex vector $\textbf{x} = \left[ x_1, x_2, \cdots, x_n \right]^T,$  $\tilde{\textbf{x}}$ is defined as
\begin{equation*}
\tilde{\textbf{x}} \triangleq \left[ \mathcal{R}\left(x_1\right), \mathcal{I}\left(x_1\right), \cdots, \mathcal{I}\left(x_n\right)\right ]^T
\end{equation*}
and for  a complex number $s$, the $\check{\left(.\right)}$ operator is defined by
\begin{equation*}
\check{\textbf{s}} \triangleq \left[\begin{array}{cc}
\mathcal{R}\left(s\right) & -\mathcal{I}\left(s\right)\\
\mathcal{I}\left(s\right) &  \mathcal{R}\left(s\right)\\
\end{array}\right].
\end{equation*}
\noindent
The $\check{\left(.\right)}$ operator can be extended to a complex $n\times n$ matrix by applying it to all the entries of it.

\section{CODE DESIGN CRITERIA}
\label{sec2}
A finite set of complex matrices is a STBC. A $n\times n$ linear STBC is obtained starting from an $n\times n$ matrix consisting of arbitrary linear combinations of $k$ complex variables and their conjugates, and letting the variables take values from complex constellations. The rate of such a code is $\frac{k}{n}$ complex symbols per channel use.
We consider Rayleigh quasi-static flat fading MIMO channel with full channel state information (CSI) at the receiver but not at the transmitter. For $2\times2$ MIMO transmission, we have
\begin{equation}
\label{Y}
\textbf{Y = HS + N}
\end{equation}
\noindent where $\textbf{S} \in \mathbb{C}^{2\times2}$ is the codeword matrix, transmitted over 2 channel uses, $\textbf{N} \in \mathbb{C}^{2\times2}$ is a complex white Gaussian noise matrix with i.i.d entries, i.e., $\sim
\mathcal{N}_{\mathbb{C}}\left(0,N_{0}\right)$ and $\textbf{H} \in \mathbb{C}^{2\times2}$ is the channel matrix with the entries assumed to be i.i.d circularly symmetric Gaussian random variables $\sim \mathcal{N}_\mathbb{C}\left(0,1\right)$. $\textbf{Y} \in \mathbb{C}^{2\times2}$ is the received matrix.

\begin{definition}\label{def1}$\left(\textbf{Code rate}\right)$ If there are $k$ independent information symbols in the codeword which are transmitted over $T$ channel uses, then, for an  $n_t\times n_r$ MIMO system, the code rate is defined as $k/T$ symbols per channel use. If $k = n_{min}T$, where $n_{min} = min\left(n_t,n_r\right)$, then the STBC is said to have $full$ $rate$.
\end{definition}

\noindent Considering ML decoding, the decoding metric that is to be minimized over all possible values of codewords $\textbf{S}$ is given by
 \begin{equation}
\label{ML}
 \textbf{M}\left(\textbf{S}\right) = \Vert \textbf{Y} - \textbf{HS} \Vert_F^2
 \end{equation}

\begin{definition}\label{def2}$\left(\textbf{Decoding complexity}\right)$
The ML decoding complexity is given  by the minimum number of symbols that  need to be jointly decoded in minimizing the decoding metric. This can never be greater than $k$, in which case, the decoding complexity is said to be of the order of $M^k$. If the decoding complexity is lesser than $M^k$, the code is said to admit simplified decoding.
\end{definition}
\begin{definition}\label{def3}$\left(\textbf{Generator matrix}\right)$ For any STBC $\textbf{S}$ that encodes $k$ information symbols, the $generator$ matrix $\textbf{G}$ is defined by the following equation
\begin{equation}
\widetilde{\textbf{vec}\left(\textbf{S}\right)} = \textbf{G} \tilde{\textbf{s}}.
\end{equation}
\noindent where $\textbf{s} \triangleq \left[ s_1, s_2,\cdots,s_k \right]^T$ is the information symbol vector
\end{definition}

The code design criteria \cite{TSC} are: (i) $Rank$ $criterion-$ To achieve maximum diversity, the codeword difference matrix $(\textbf{X} - \hat{\textbf{X}})$ must be full rank for all possible pairs of codewords and the diversity gain is given by $n_tn_r,$ (ii) $Determinant$ $criterion-$ For a full ranked STBC, the minimum determinant $\delta_{min}$, defined as
\begin{equation}
 \delta_{min} \triangleq \min_{\textbf{X} \neq \hat{\textbf{X}}} det\left[\left(\textbf{X}-\hat{\textbf{X}}\right)\left(\textbf{X}-\hat{\textbf{X}}\right)^{H}\right]
\end{equation}
should be maximized. The coding gain is given by $\left(\delta_{min}\right)^{1/n_t}$, with $n_t$ being the number of transmit antennas.

%

For the $2\times2$ MIMO system, the target is to design a code that is full-rate, i.e transmits 2 complex symbols per channel use,  has full-diversity, maximum coding gain and allows low ML decoding complexity.

\section{ THE PROPOSED STBC }
\label{sec3}
In this section, we present our STBC for  $2\times 2$ MIMO system. The design is based on the class of codes called co-ordinate interleaved orthogonal designs (CIODs), which was studied in \cite{ZS} in connection with the general class of single-symbol decodable codes and, specifically for 2 transmit antennas, is as follows.
\begin{definition}
 The CIOD for $2$ transmit antennas \cite{ZS} is \\
\begin{equation}
\label{ciod}
\textbf{X}(s_1,s_2) = \left[\begin{array}{cc}
            s_{1I}+js_{2Q} & 0 \\
            0  & s_{2I}+js_{1Q}\\
           \end{array}\right]
\end{equation}
where $s_i \in \mathbb{C}, i = 1,2$ are the information symbols and $s_{iI}$ and $s_{iQ}$ are the in-phase (real) and quadrature-phase (imaginary) components of $s_i,$ respectively. Notice that in order to make the above STBC full rank, the signal constellation $\mathcal{A}$ from which the symbols $s_i$ are chosen should be such that the real part (imaginary part, resp.) of any signal point in $\mathcal{A}$ is not equal to the real part (imaginary part, resp.) of any other signal point in $\mathcal{A}$ \cite{ZS}. So if QAM constellations are chosen, they have to be rotated. The optimum angle of rotation has been found in \cite{ZS} to be $\frac{1}{2} tan^{-1}2$ degrees and this maximizes the diversity and coding gain. We denote this angle by $\theta_g.$
\end{definition}

The proposed  $2 \times 2$ STBC $\mathbf{S}$ is given by
\begin{equation}
 \textbf{S}(x_1,x_2,x_3,x_4) = \textbf{X}\left(s_1,s_2\right) + e^{j\theta}\textbf{X}\left(s_3,s_4\right)\textbf{P}
\end{equation}
where
\begin{itemize}
\item The four symbols $s_1,s_2,s_3$ and $s_4 \in \mathcal{A}$, where $\mathcal{A}$ is a $\theta_g$ degrees rotated version of a regular QAM signal set, denoted by $\mathcal{A}_q$ which is a finite subset of the integer lattice, and $x_1,x_2,x_3,x_4 \in \mathcal{A}_q.$ To be precise, $s_i=e^{\theta_g}x_i,~~~i=1,2,3,4.$

\item $\textbf{P}$ is a permutation matrix designed to make the STBC full rate and is given by $
 \textbf{P} = \left[\begin{array}{cc}
0 & 1 \\
1 & 0 \\
\end{array}\right].$
\item The choice of $\theta$ in the above expression should be such that the diversity and coding gain are maximized. A computer search was done for $\theta$ in the range $\left[ 0 , \pi/2 \right]$. The optimum value of $\theta$ was found out to be $\pi/4$.
\end{itemize}

Explicitly, our code matrix is

$\textbf{S}(x_1,x_2,x_3,x_4)  =$
\begin{equation}
   \left[\begin{array}{rr}
 s_{1I}+js_{2Q} & e^{j\pi/4}(s_{3I}+js_{4Q})  \\
 e^{j\pi/4}(s_{4I}+js_{3Q}) & s_{2I}+js_{1Q}
\end{array}\right] \\
\end{equation}
The minimum determinant for our code when the symbols are chosen from QAM constellations is $3.2$, the same as that of the Golden code, which will be proved in the next section.

The generator matrix for our STBC, corresponding to the symbols $s_i$, is as follows:
\begin{equation}\label{gen}
G = \left[\begin{array}{cccccccc}
      1 & 0 & 0 & 0 & 0 & 0 & 0 & 0 \\
      0 & 0 & 0 & 1 & 0 & 0 & 0 & 0 \\
      0 & 0 & 0 & 0 & 0 & -\frac{1}{\sqrt{2}} & \frac{1}{\sqrt{2}} & 0\\
      0 & 0 & 0 & 0 & 0 & \frac{1}{\sqrt{2}} & \frac{1}{\sqrt{2}} & 0\\
      0 & 0 & 0 & 0 & \frac{1}{\sqrt{2}} & 0 & 0 & -\frac{1}{\sqrt{2}}\\
      0 & 0 & 0 & 0 & \frac{1}{\sqrt{2}} & 0 & 0 & \frac{1}{\sqrt{2}}\\
      0 & 0 & 1 & 0 & 0 & 0 & 0 & 0\\
      0 & 1 & 0 & 0 & 0 & 0 & 0 & 0 \\
      \end{array}\right]
\end{equation}
It is easy to see that this generator matrix is orthonormal. In \cite{ZLW}, it was shown that a necessary and sufficient condition for an STBC to be \emph{Information lossless} is that its generator matrix should be unitary. Hence, our STBC has the \emph{Information losslessness} property.

\section{NVD PROPERTY AND THE DMG OPTIMALITY}
\label{sec4}
In this section it is shown that the proposed code has the non-vanishing determinant (NVD) property \cite{BRV}, which in conjunction with full-rateness means that our code is  DMG tradeoff optimal \cite{PVK}.

 The determinant of the codeword matrix $\textbf{S}$ can be written as
\begin{equation*}
det(\textbf{S}) = (s_{1I}+js_{2Q})(s_{2I}+s_{1Q}) - j[(s_{3I}+js_{4Q})(s_{4I}+s_{3Q})].
\end{equation*}
Using $s_{iI} = (s_i+s_i^*)/2$ and $js_{iQ} = (s_i-s_i^*)/2$ in the equation above, we get,
{\footnotesize
\begin{eqnarray*}
4det(\textbf{S}) & = &(s_1+s_1^*+s_2-s_2^*)(s_2+s_2^*+s_1-s_1^*) {}
\nonumber\\
&&{}-j[(s_3+s_3^*+s_4-s_4^*)(s_4+s_4^*+s_3-s_3^*)]\\
 & = &\big((s_1+s_2)+(s_1-s_2)^*\big)\big((s_1+s_2)-(s_1-s_2)^*\big){}
\nonumber\\
&&{}-j[\big((s_3+s_4)+(s_3-s_4)^*\big)\big((s_3+s_4)-(s_3-s_4)^*\big)].
\end{eqnarray*}
}
Since $s_i = e^{j\theta_g}x_i, i = 1,2,3,4$, with $s_i \in \mathcal{A}$, $x_i \in \mathcal{A}_q$, a subset of $\mathbb{Z}[i]$,  defining $A \triangleq (x_1+x_2)$, $B \triangleq (x_1-x_2)^*$, $C \triangleq (x_3+x_4)$ and $D \triangleq (x_3-x_4)^*$, with $A,B,C$ and $D \in \mathbb{Z}[i]$, we get
\begin{eqnarray*}
4Det(\textbf{S})& = & (e^{j\theta_g}A+e^{-j\theta_g}B)(e^{j\theta_g}A-e^{-j\theta_g}B){}
\nonumber\\
&&{}-j[(e^{j\theta_g}C+e^{-j\theta_g}D)(e^{j\theta_g}C-e^{-j\theta_g}D)]\\
& = & e^{j2\theta_g}A^2-e^{-j2\theta_g}B^2 - j[e^{j2\theta_g}C^2-e^{-j2\theta_g}D^2].
\end{eqnarray*}
Since $e^{j2\theta_g} = cos(2\theta_g)+sin(2\theta_g)=(1+2j)/\sqrt{5}$, we get
\begin{equation}\label{mod}
4\sqrt{5}Det(\textbf{S}) =  (1+2j)(A^2-jC^2) - (1-2j)(B^2-jD^2).
\end{equation}
For the determinant of $\textbf{S}$ to be 0, we must have
\begin{eqnarray*}
 (1+2j)(A^2-jC^2)& = &(1-2j)(B^2-jD^2) \\
\Rightarrow (1+2j)^2(A^2-jC^2)& = &5(B^2-jD^2).
\end{eqnarray*}
The above can be written as
\begin{equation}\label{ABC}
 A_1^2-jC_1^2 = 5(B^2-jD^2)
\end{equation}
where $A_1 = (1+2j)A,C_1 = (1+2j)C$ and clearly $A_1,C_1 \in \mathbb{Z}[i]$. It has been shown in \cite{DV} that  \eqref{ABC} holds only when $A_1 = B = C_1 = D = 0$, i.e., only when $x_1 = x_2 =x_3 = x_4 =0$. This means that the determinant of the codeword difference matrix is 0 only when the codeword difference  matrix is itself 0. So, for any distinct pair of codewords, the codeword difference matrix is always full rank for any constellation which is a subset of $\mathbb{Z}[i]$. Also, the minimum value of the modulus of R.H.S of \eqref{mod} can be seen to be $4$. So, $\vert Det(\textbf{S}) \vert \geq 1/\sqrt{5}$. In particular, when the constellation chosen is the standard QAM constellation, the difference between any two signal points is a multiple of 2. Hence, for such constellations, $\vert Det(\textbf{S-S}^\prime) \vert \geq 4/\sqrt{5}$, where $\textbf{S}$ and $\textbf{S}^\prime$ are distinct codewords. The minimum determinant is consequently 16/5. This means that the proposed codes has the non-vanishing determinant (NVD) property \cite{BRV}. In \cite{PVK}, it was shown that full-rate codes which satisfy the non-vanishing determinant property achieve the optimal DMG tradeoff. So, our proposed STBC is DMG tradeoff optimal.
\section{DECODING COMPLEXITY}
\label{sec5}
The decoding complexity of the proposed code is of the order of $M^3$. This is due to the fact that conditionally given the symbols $x_3$ and $x_4$, the symbols $x_1$ and $x_2$ can be decoded independently. This can be proved as follows. Writing the STBC in terms of its weight matrices/dispersion matrices $A_i,~~i=1,2,\cdots,8,$   \cite{ZS}, we have
\begin{eqnarray*}
 {\mathbf S} & = &\sum_{m=1}^{4}\underbrace{x_{mI}A_{2m-1} + x_{mQ}A_{2m}}_{\textbf{T}_m} = S_1 + S_2
\end{eqnarray*}
\noindent where
\begin{equation*}
 S_1 = \sum_{m=1}^{2}x_{mI}A_{2m-1} + x_{mQ}A_{2m}
\end{equation*}
\noindent and
\begin{equation*}
 S_2 = \sum_{m=3}^{4}x_{mI}A_{2m-1} + x_{mQ}A_{2m}.
\end{equation*}
For our code, we have
\begin{eqnarray*}
 A_{1}& = &\left[\begin{array}{rr}
              cos\theta_g & 0 \\
              0 & jsin\theta_g \\
 \end{array}\right ]; ~~~~~~~ A_{2} =  \left[\begin{array}{rr}
              -sin\theta_g & 0 \\
              0 & jcos\theta_g \\
 \end{array}\right]\\
A_{3}& = &\left[\begin{array}{rr}
              jsin\theta_g & 0 \\
              0 & cos\theta_g \\
 \end{array}\right ]; ~~~~~~~ A_{4} =  \left[\begin{array}{rr}
              jcos\theta_g & 0 \\
              0 & -sin\theta_g \\
 \end{array}\right]\\
A_{5}& = &e^{j\pi/4}\left[\begin{array}{rr}
              0 & cos\theta_g   \\
              jsin\theta_g & 0 \\
 \end{array}\right ]\\
 A_{6}& = & e^{j\pi/4}\left[\begin{array}{rr}
              0 & -sin\theta_g \\
              jcos\theta_g & 0  \\
 \end{array}\right]
\end{eqnarray*}
\begin{eqnarray*}
A_{7}& = & e^{j\pi/4}\left[\begin{array}{rr}
              0 & jsin\theta_g  \\
              cos\theta_g & 0  \\
 \end{array}\right ]\\
 A_{8}& = & e^{j\pi/4}\left[\begin{array}{rr}
              0 & jcos\theta_g  \\
              -sin\theta_g & 0  \\
 \end{array}\right].
\end{eqnarray*}
The ML decoding metric in \eqref{ML} can be written as
\begin{eqnarray*}
M\left(S\right) & = & tr\left[\left(Y-HS\right)\left(Y-HS\right)^H\right]\\
& = & tr\left[\left(Y-HS_1-HS_2\right)\left(Y-HS_1-HS_2\right)^H\right]\\
& = & tr\left[\left(Y-HS_1\right)\left(Y-HS_1\right)^H\right] {}
                                                       \nonumber\\
&&{}-tr\left[HS_2\left(Y-HS_1\right)^H\right]{}
                                          \nonumber\\
&&{}-tr\left[\left(Y-HS_1\right)\left(HS_2\right)^H\right] {}
\nonumber\\
&&{}+ tr\left[HS_2\left(HS_2\right)^H\right].
\end{eqnarray*}
It can be verified that the following hold true for $l,m \in \left[1,4\right]$
\begin{equation*}
 A_mA_l^H + A_lA_m^H = 0   \left\{ \begin{array}{ll}
\forall l \neq m, m+1, & \textrm{if m is odd}\\
\forall l \neq m, m-1, & \textrm{if m is even}.\\
\end{array} \right.
\end{equation*}
\noindent From \cite{ZS}, we obtain
{\footnotesize
\begin{eqnarray*}
 tr\left[\left(Y-HS_1\right)\left(Y-HS_1\right)^H\right] =
 \sum_{m=1}^{2}\Vert Y - HT_m\Vert_F^2-tr\left(YY^H\right)
\end{eqnarray*}}
\noindent and hence,
\begin{eqnarray*}
M\left(S\right) & = & \sum_{m=1}^{2}\Vert Y - HT_m\Vert_F^2-tr\left(YY^H\right) {}
                                                                  \nonumber\\
&& {}+tr\left[HS_2\left(HS_1\right)^H\right]+tr\left[HS_1\left(HS_2 \right)^H\right]
 \nonumber\\
&& {}-tr\left[HS_2 Y^H\right]-tr\left[Y\left(HS_2 \right)^H\right]{}
\nonumber\\
&&{}+tr\left[HS_2\left(HS_2\right)^H\right]
\end{eqnarray*}
\begin{eqnarray*}
& = & \sum_{m=1}^{2}\Vert Y - HT_m\Vert_F^2 +\sum_{m=1}^{2}tr\left[HS_2\left(HT_m\right)^H\right]{}
\nonumber\\
&& {}+\sum_{m=1}^{2}tr\left[HT_m\left(HS_2\right)^H\right] +\Vert Y-HS_2\Vert_F^2{}
\nonumber\\
&&{}-2tr(YY^H).
\end{eqnarray*}
Hence, when $S_2$ is given, i.e, symbols $x_3$ and $x_4$ are given, the ML metric can be decomposed as
\begin{equation}
 M\left(S\right) = \sum_{m=1}^{2}M\left(x_m\right) + M_c
\end{equation}
\noindent with $M_c = \Vert Y-HS_2\Vert_F^2{} -2tr(YY^H) $ and $M(x_m)$ being a function of symbol $x_m$ alone. Thus decoding can be done as follows: choose the pair $\left(x_3,x_4\right)$ and then, in parallel, decode $x_1$ and $x_2$ so as to minimize the ML decoding metric. With this approach, there are $2M^3$ values of the decoding metric that need to be computed in the worst case. So, the decoding complexity is of the order of $M^3$.

\section{SIMPLIFIED DECODING USING SPHERE DECODER}
\label{sec6}
In this section, it is shows that sphere decoding can be used to achieve the decoding complexity of $M^3$.
It can be shown that \eqref{Y} can be written as
\begin{equation}\label{eqmod}
 \widetilde{vec(\textbf{Y})} = \textbf{H}_{eq}\tilde{\textbf{s}} + \widetilde{vec(\textbf{N})}
\end{equation}
\noindent where $\textbf{H}_{eq} \in \mathbb{R}^{8\times8}$ is given by
\begin{equation}
 \textbf{H}_{eq} = \left(\textbf{I}_2 \otimes \check{\textbf{H}}\right)\textbf{G}
\end{equation}
with $\textbf{G} \in \mathbb{R}^{8\times8}$ being the generator matrix as in \eqref{gen} and
\begin{equation*}
\tilde{\textbf{s}} \triangleq \left[\mathcal{R}(s_1),\mathcal{I}(s_1),\cdots,\mathcal{R}(s_4),\mathcal{I}(s_4)\right]^T
\end{equation*}
\noindent with $s_i,i=1,\cdots,4$ drawn from $\mathcal{A}$, which is a rotation of the regular QAM constellation $\mathcal{A}_q$. Let
\begin{center}
$\textbf{x}_q \triangleq [ x_1, x_2, x_3, x_4 ]^T$
\end{center}
\noindent Then,
\begin{equation*}
 \tilde{\textbf{s}} = \textbf{F}\tilde{\textbf{x}}_q.
\end{equation*}
\noindent where $\textbf{F} \in \mathbb{R}^{8\times8}$ is $diag[\textbf{J},\textbf{J},\textbf{J},\textbf{J}]$ with $\textbf{J}$ being a rotation matrix and is defined as follows
\begin{equation*}
 \textbf{J} \triangleq \left[\begin{array}{cc}
                     cos(\theta_g) & -sin(\theta_g)\\
                     sin(\theta_g) &  cos(\theta_g)\\
              \end{array}\right].
\end{equation*}
So, \eqref{eqmod} can be written as
\begin{equation}
\widetilde{vec(\textbf{Y})} = \textbf{H}_{eq}^\prime\tilde{\textbf{x}_q} + \widetilde{vec(\textbf{N})}
\end{equation}
\noindent where $\textbf{H}_{eq}^\prime = \textbf{H}_{eq}\textbf{F}$. 
Using this equivalent model, the ML decoding metric can be written as
\begin{equation}
 \textbf{M}\left(\tilde{\textbf{x}_q}\right) = \Vert \widetilde{vec\left(\textbf{Y}\right)} - \textbf{H}_{eq}^\prime\tilde{\textbf{x}_q}\Vert^2
\end{equation}
On obtaining the QR decomposition of $\textbf{H}_{eq}^\prime$, we get $\textbf{H}_{eq}^\prime $= $\textbf{QR}$, where  $\textbf{Q} \in \mathbb{R}^{8\times8}$ is an orthonormal matrix and  $\textbf{R} \in \mathbb{R}^{8\times8}$ is an upper triangular matrix. The ML decoding metric now can be written as
\begin{equation}
 \textbf{M}(\tilde{\textbf{x}_q}) = \Vert \textbf{Q}^T\widetilde{\textbf{vec(Y)}} - \textbf{R}\tilde{\textbf{x}_q}\Vert^2
\end{equation}
If $\textbf{H}_{eq}^\prime \triangleq [ \textbf{h}_1 \ \textbf{h}_2 \cdots \textbf{h}_{8} ]$, where $\textbf{h}_i, i = 1,2,\cdots,8$ are column vectors, then $\textbf{Q}$ and $\textbf{R}$ have the general form obtained by $Gram-Schmidt$ process as shown below
\begin{equation*}
 \textbf{Q} = [ \textbf{q}_1\ \textbf{q}_2 \ \textbf{q}_3 \cdots \textbf{q}_{8} ]
\end{equation*}
where $\textbf{q}_i, i = 1,2,\cdots,8$ are column vectors, and
\begin{equation*}
 \textbf{R} = \left[\begin{array}{ccccc}
\Vert \textbf{r}_1 \Vert & \langle \textbf{h}_2,\textbf{q}_1 \rangle & \langle \textbf{h}_3,\textbf{q}_1 \rangle & \ldots &  \langle \textbf{h}_{8},\textbf{q}_1 \rangle\\
0 & \Vert \textbf{r}_2 \Vert & \langle \textbf{h}_3,\textbf{q}_2 \rangle & \ldots & \langle \textbf{h}_{8},\textbf{q}_2 \rangle\\
0 & 0 &  \Vert \textbf{r}_3 \Vert & \ldots & \langle \textbf{h}_{8},\textbf{q}_3 \rangle\\
\vdots & \vdots & \vdots & \ddots & \vdots\\
0 & 0 & 0 & \ldots & \Vert \textbf{r}_{8} \Vert\\
\end{array}\right]
\end{equation*}
\noindent where $\textbf{r}_1 = \textbf{h}_1$, $\ \textbf{q}_1 = \frac{\textbf{r}_1}{\Vert \textbf{r}_1 \Vert},$ $\textbf{r}_i = \textbf{h}_i - \sum_{j=1}^{i-1}\langle \textbf{h}_i,\textbf{q}_j \rangle \textbf{q}_j,$ $\ \textbf{q}_i = \frac{\textbf{r}_i}{\Vert \textbf{r}_i \Vert},\ i = 2,\cdots,8.$

It can be shown by direct computation that $\textbf{R}$ has the following structure
\begin{equation}
\label{formofR} \left[ \begin{array}{cccccccc}
a & a & 0 & 0 & a & a & a & a \\ 0 & a & 0 & 0 & a & a & a & a \\ 0 & 0 & a & a & a & a & a & a \\
0 & 0 & 0 & a & a & a & a & a \\
0 & 0 & 0 & 0 & a & a & a & a \\
0 & 0 & 0 & 0 & 0 & a & a & a \\
0 & 0 & 0 & 0 & 0 & 0 & a & a \\
0 & 0 & 0 & 0 & 0 & 0 & 0 & a \end{array} \right] \end{equation} where $a$ stands for a possibly non-zero entry.

The structure of the matrix ${\textbf{R}}$ allows us to perform a 4 dimensional real sphere decoding (SD) \cite{ViB} to find the partial vector $[ \mathcal{R}(x_3), \mathcal{I}(x_3),\mathcal{R}(x_4),\mathcal{I}(x_4) ]^T$ and hence obtain the symbols $x_3$ and $x_4$. Having found these, $x_1$ and $x_2$ can be decoded independently. Observe that the real and imaginary parts of symbol $x_1$ are entangled with one another because of constellation rotation but are independent of the real and imaginary parts of $x_2$ when $x_3$ and $x_4$ are conditionally given.

Having found the partial vector $[\mathcal{R}(x_3),\mathcal{I}(x_3),\mathcal{R}(x_4),\mathcal{I}(x_4) ]^T$, we proceed to find the rest of the symbols as follows. We do two parallel 2 dimensional real search to decode the symbols $x_1$ and $x_2$. So, overall, the worst case decoding complexity of the proposed STBC is 2$M^3$. This is due to the fact that
\begin{enumerate}
 \item A 4 dimensional real SD requires $M^2$ metric computations in the worst possible case.
 \item Two parallel 2 dimensional real SD require $2M$ metric computations in the worst case.
\end{enumerate}
This decoding complexity is the same as that achieved by the HTW-PGA code and the Sezginer-Sari code.

Though it has not been mentioned anywhere to the best of our knowledge, the ML decoding complexity of the Golden code, Dayal-Varanasi code and the trace-orthogonal cyclotomic code is also $2M^3$ for square QAM constellations. This follows from the structure of the $\textbf{R}$ matrices for these codes which are counterparts of the one in \eqref{formofR}. The $\textbf{R}$ matrices of these codes are similar in structure and as shown below:

\begin{equation*}
\textbf{R}=
\left[
\begin{array}{cccccccc}
a & 0 & a & 0 & a & a & a & a \\
0 & a & 0 & a & a & a & a & a \\
0 & 0 & a & 0 & a & a & a & a \\
0 & 0 & 0 & a & a & a & a & a \\
0 & 0 & 0 & 0 & a & a & a & a \\
0 & 0 & 0 & 0 & 0 & a & a & a \\
0 & 0 & 0 & 0 & 0 & 0 & a & a \\
0 & 0 & 0 & 0 & 0 & 0 & 0 & a
\end{array}
\right]
\end{equation*}

Table \ref{table1} presents the comparison of the known full-rate, full-diversity $2\times 2$ codes in terms of their ML decoding complexity and the coding gain.

\section{SIMULATION RESULTS}
\label{sec7}
\begin{figure}
\centering
\includegraphics[width=3.5in, height=3.5in]{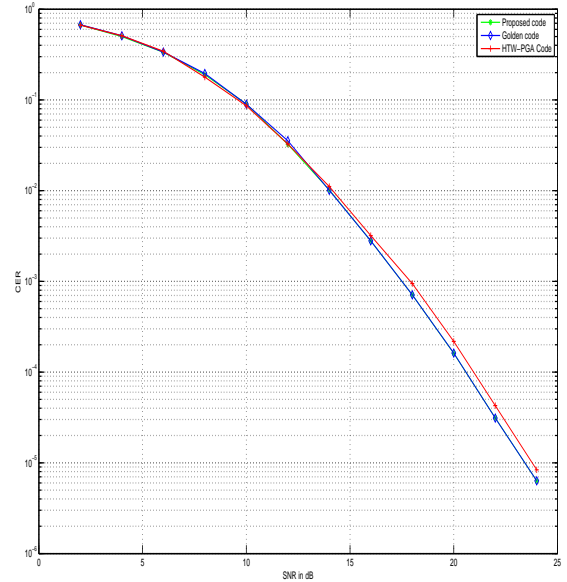}
\caption{CER PERFORMANCE FOR 4-QAM}
\label{4qam}
\end{figure}

\begin{figure}
\centering
\includegraphics[width=3.5in, height=3.5in]{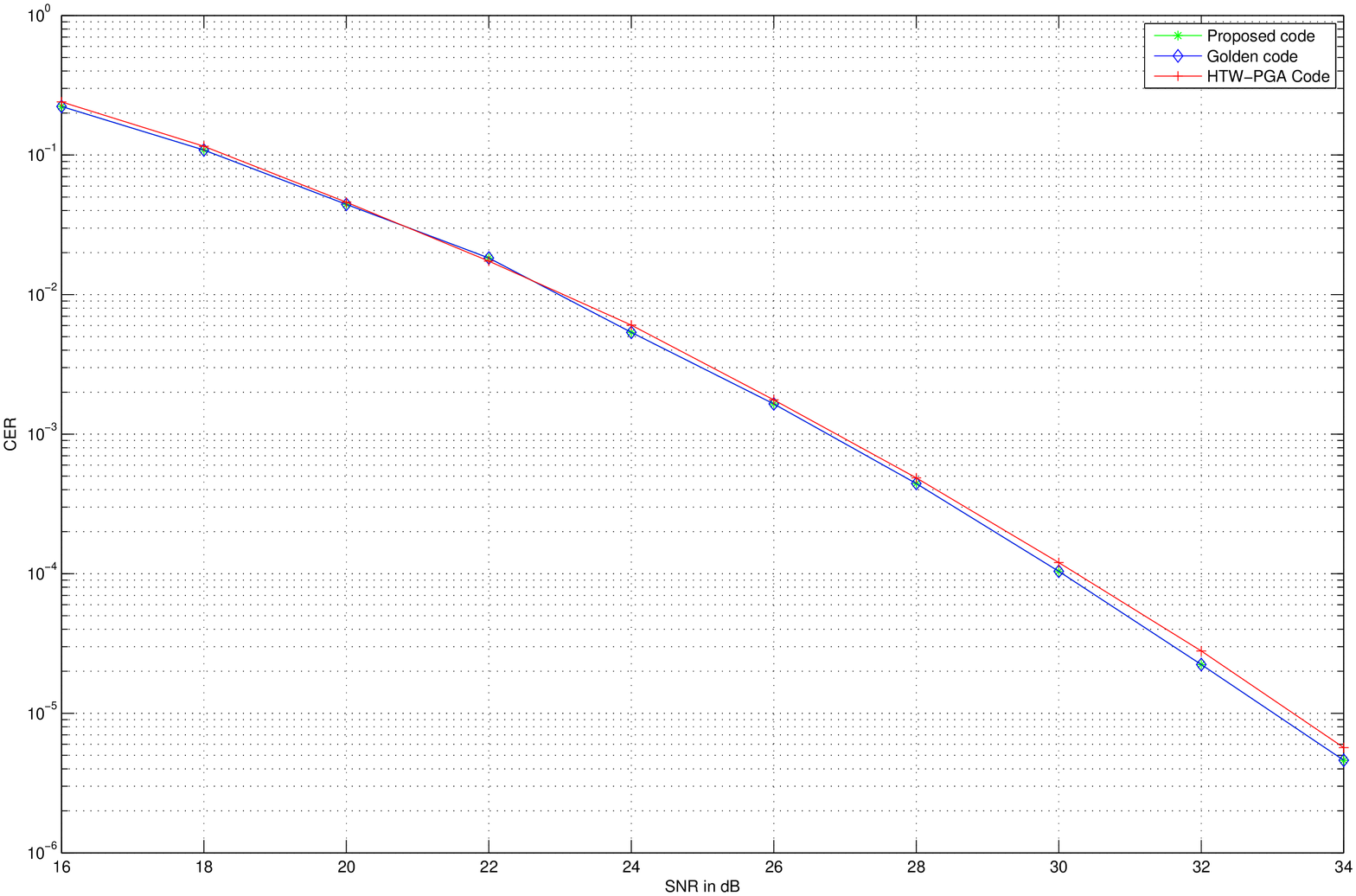}
\caption{CER PERFORMANCE FOR 16-QAM}
\label{16qam}
\end{figure}

\begin{figure}
\centering
\includegraphics[width=3.5in, height=3.5in]{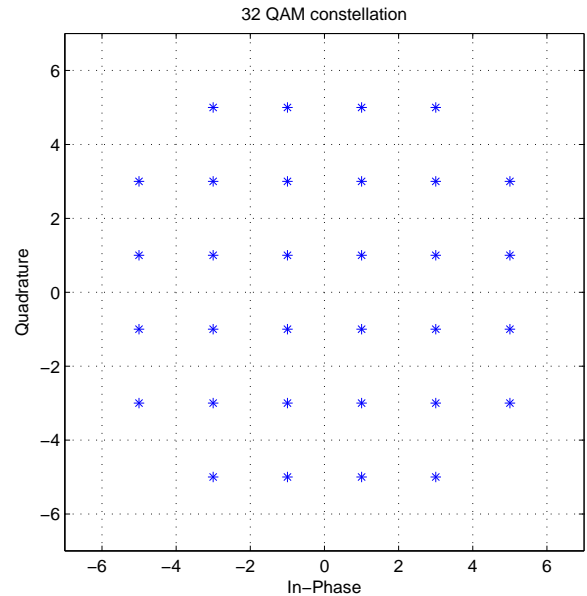}
\caption{AN EXAMPLE OF A NON-RECTANGULAR QAM CONSTELLATION}
\label{fig2}
\end{figure}

Fig \ref{4qam} shows the codeword error performance plots for the Golden code, the proposed STBC and the HTW-PGA code for the 4-QAM constellation. The performance of the proposed code is the same as that of the Golden code. The HTW-PGA code performs slightly worse due to its lower coding gain. Fig \ref{16qam}, which is a plot of the CER performance for 16-QAM, also highlights these aspects. Table \ref{table1} gives a comparison between  the well known full-rate, full-diversity codes for $2\times2$ MIMO.

\begin{table*}
\begin{center}
\begin{tabular}{|c|c|c|c|c|} \hline
     & Min det  & \multicolumn{3}{c|}{ ML Decoding complexity}  \\ \cline{3-5}
 Code & for QAM  & Square QAM &  Rectangular QAM & Non-rectangular  \\
      &          &            &  $M=M_1\times M_2$    &  QAM  \\ \hline
Yo-Wornell\cite{YW} &  0.8000  & 2$M^3$ & $M^2(M_1^2 + M_2^2)$ &  $M^4$ \\ \hline
Dayal-Varanasi code\cite{DV} &  3.2000  & 2$M^3$ & $M^2(M_1^2 + M_2^2)$ &  $M^4$ \\ \hline
Golden code \cite{BRV}&  3.2000  & 2$M^3$ & $M^2(M_1^2 + M_2^2)$ &  $M^4$ \\ \hline
Trace-orthonormal cyclotomic code \cite{ZLW} &  3.2000 & 2$M^3$ & $M^2(M_1^2 + M_2^2)$ &  $M^4$ \\ \hline
HTW-PGA code \cite{PGA}&  2.2857 & 4$M^2\sqrt{M}$ & 2$M^2(M_1+M_2)$ &  $2M^3$ \\ \hline
Sezginer-Sari code \cite{SS}&  2.0000 & 4$M^2\sqrt{M}$  & 2$M^2(M_1+M_2)$ &  $2M^3$ \\ \hline
The proposed code &  3.2000 & $2M^3$ & $2M^3$  & $2M^3$ \\ \hline
\end{tabular}
\end{center}
\caption{COMPARISION BETWEEN THE MINIMUM DETERMINANT AND DECODING COMPLEXITY OF SOME WELL KNOWN FULL-RATE $2\times2$ STBCs}
\label{table1}
\end{table*}

\section{CONCLUDING REMARKS}
\label{sec8}
In this paper, we have presented a full-rate STBC for $2\times2$ MIMO systems which matches the best known codes for such systems in terms of error performance, while at the same time, enjoys  simplified-decoding complexity that the codes presented in \cite{PGA} and \cite{SS} do. Recently, a Rate-1 STBC, based on scaled repetition and rotation of the Alamouti code, was proposed \cite{FW}. This code was shown to have a hard-decision performance which was only slightly worse than that of the Golden code for a spectral efficiency of $4b/s/Hz$, but the complexity was significantly lower.

\section*{ACKNOWLEDGEMENT}
This work was partly supported by the DRDO-IISc program on Advanced Research in Mathematical Engineering.



\begin{thebibliography}{1}

\bibitem{SMA}
S. M. Alamouti, ``A simple transmit diversity technique for wireless communications'', \emph{IEEE J. Sel. Areas Commun.}, vol.\ 16, no.\ 8, pp. 1451-1458, October 1998.

\bibitem{TJC}
V. Tarokh, H. Jafarkhani and A. R. Calderbank, ``Space-time block codes from orthogonal designs'', \emph{IEEE Trans. Inf. Theory}, vol. \ 45, no.\ 5, pp. 1456-1467, July 1999.

\bibitem{BRV}
J. C. Belfiore, G. Rekaya and E. Viterbo, ``The Golden Code: A $2\times2$ full rate space-time code with non-vanishing determinants,'' \emph{IEEE Trans. Inf. Theory}, vol. \ 51, no. \ 4, pp. \ 1432-1436, April 2005.

\bibitem{DV}
P. Dayal, M. K. Varanasi, "An optimal two transmit antenna space-time code and its stacked extensions,"  \emph{IEEE Trans. Inf. Theory}, vol.\ 51, no.\ 12, pp. 4348-4355, Dec. 2005.

\bibitem{ZLW}
Jian-Kang Zhang, Jing Liu, Kon Max Wong, "Trace-Orthonormal Full-Diversity Cyclotomic Space–Time Codes," \emph{IEEE Transactions on  Signal Processing} , vol. \ 55, no.\ 2, pp.618-630, Feb 2007.

\bibitem{HTW}
A. Hottinen, O. Tirkkonen and R. Wichman, "Multi-antenna Transceiver Techniques for 3G and Beyound," WILEY publisher, UK.

\bibitem{PGA}
J. Paredes,  A.B. Gershman and  M. Gharavi-Alkhansari, ''A $2\times2$ Space-Time Code with Non-Vanishing Determinants and Fast Maximum Likelihood Decoding," in Proc \emph{IEEE International Conference on Acoustics, Speech and Signal Processing(ICASSP 2007),} vol.\ 2, pp.877-880, April 2007.

\bibitem{SS}
S. Sezginer and H. Sari, ``A full rate full-diversity $2\times2$ space-time code for mobile Wimax Systems,'' in Proc. \emph{ IEEE International Conference on Signal Processing and Communications}, Dubai, July 2007.

\bibitem{BHV}
E. Biglieri, Y. Hong and E. Viterbo, ''On Fast-Decodable Space-Time Block Codes``, submitted to \emph{IEEE Trans. Inf. Theory}.

\bibitem{PVK}
P. Elia, K. R. Kumar, S. A. Pawar, P. V. Kumar and H. Lu, ''Explicit construction of space-time block codes: Achieving the diversity-multiplexing gain tradeoff``, \emph{IEEE Trans. Inf. Theory}, vol. \ 52, pp. 3869-3884, Sept. 2006.

\bibitem{ZS}
Zafar Ali Khan, Md., and B. Sundar Rajan, ``Single Symbol Maximum Likelihood Decodable Linear STBCs'', \emph{IEEE Trans. Inf. Theory}, vol.\ 52, No.\ 5, pp. 2062-2091, May 2006.

\bibitem{TSC}
 V.Tarokh, N.Seshadri and A.R Calderbank,"Space time codes for high date rate wireless communication : performance criterion and code construction'',
\emph{IEEE Trans. Inf. Theory}, vol.\ 44,  pp. 744 - 765, 1998.

\bibitem{ViB}
Emanuele Viterbo and Joseph Boutros, ``Universal lattice code decoder for fading channels'', \emph{IEEE Trans. Inf. Theory}, vol.\ 45, No.\ 5, pp. 1639-1642, July 1999.

\bibitem{YW}
H. Yao and G. W. Wornell, ``Achieving the full MIMO diversity-multiplexing frontier with rotation-based space-time codes,'' in \emph{Proc. Allerton Conf. on Comm. Control and Comput.,} Monticello, IL, Oct. 2003.

\bibitem{FW}
F. M. J. Willems, ``Rotated and Scaled Alamouti Coding'', arXiv:0802.0580(cs.IT),  February 5, 2008


\end{thebibliography}
\end{document}